\documentclass[11pt,twoside]{article}
\usepackage{asp2010}

\resetcounters

\begin{document}

\title{The CfAO's Astronomy Course in COSMOS: Curriculum Design,
  Rationale, and Application}

\author{Kathy~Cooksey,$^1$ Scott~Seagroves,$^2$ Jason~Porter,$^3$ Lynne~Raschke,$^4$ Scott~Severson,$^5$ and Sasha~Hinkley$^6$
  \affil{$^1$Massachusetts~Institute~of~Technology, 
    Kavli~Institute~for~Astrophysics~and~Space~Research, 
    77~Massachusetts~Avenue,
    37--611, Cambridge,~MA~02139}
  \affil{$^2$Institute~for~Scientist~and~Engineer~Educators, 
    Center~for~Adaptive~Optics, UC~Santa~Cruz, 1156~High~Street,
    Santa~Cruz,~CA~95064} 
  \affil{$^3$College~of~Optometry, University~of~Houston,
    4901~Calhoun~Road, Room~2149, Houston,~TX~77204}
  \affil{$^4$School~of~Sciences, The~College~of~St.~Scholastica, 
    1200~Kenwood~Avenue, Office~S~2135C, Duluth,~MN~55811}
  \affil{$^5$Department~of~Physics~\&~Astronomy,
    Sonoma~State~University, 300L~Darwin~Hall, Rohnert~Park,~CA~94928}  
  \affil{$^6$Department~of~Astronomy,
    California~Institute~of~Technology,
    1200~East~California~Boulevard, Mail~Code~249--17, Pasadena,~CA~91125}
}

\begin{abstract} From 2001 to 2007, COSMOS provided a teaching and
  outreach venue for the Center for Adaptive Optics Professional
  Development Program (CfAO PDP). COSMOS is a four-week residential
  mathematics and science summer program for high-school students
  organized by the University of California on four of its
  campuses. Two topical science courses comprised each COSMOS cluster.
  An astronomy course has always formed a basis for the CfAO
  PDP-affiliated cluster. The course included a variety of
  pedagogical techniques to address a diversity of learners and
  goals. We outline the astronomy course---lectures, activities,
  etc.---and provide the rationale for what was taught, how it was
  taught, and when it was taught.
\end{abstract}

\section{Introducing COSMOS and the CfAO's Course Cluster}\label{sec.intro}

The California State Summer School for Mathematics and Science
(COSMOS) is a month-long residential academic and enrichment
experience for high school students. The University of California,
Santa Cruz (UCSC) is one of the four UC campuses that host a
program.\footnote{\url{http://epc.ucsc.edu/cosmos/}} For four weeks,
COSMOS students divide their time between program-wide events,
both academic and recreational, and their course ``clusters.''  A cluster is
a pairing of two math or science-based courses, such as astronomy and
vision science. While there are COSMOS program-wide academic activities
(lectures on current science research, career panels, etc.), the
majority of academic time is dedicated to the cluster. Typically
students spend about six hours per day in cluster activities. For a
wider perspective on COSMOS at UCSC, see \citet{andreasen-etal-epc}.

Each cluster has a teacher fellow (TF), two or more instructors (at
least one lead instructor per course), and a similar number of
teaching assistants. In addition, the students' social and other
activities are chaperoned and organized by COSMOS resident
assistants. The teacher fellow is a local high-school teacher and is
the one instructor who attends all academic events with the
students. The TF forms a bridge between the two courses by helping the
students synthesize the content from the courses and the COSMOS-wide
lectures as well as communicating with the instructors about the mood
of the students. The TF also teaches an approximately
six-hour-per-week section on ``Transferable Skills,'' where the
students learn about, e.g., spreadsheets and preparing presentations,
which are beneficial to both courses. Key to a successful cluster is
the TF's experience with high school students---a valuable resource
for the other instructors.  The instructors are often UCSC faculty or
doctoral students, and they are in charge of teaching lectures, labs,
etc., to the whole cluster.

The Cluster 7\footnote{The CfAO's cluster was initially Cluster 10 but
  has been Cluster 7 for most of the time. Hence, we consider Cluster
  7 synonymous with the CfAO's cluster.}  teaching assistants were
more than aides to the instructors. They were leaders of small-group
research projects; therefore, we refer to them as project
advisors. All COSMOS students participate in a research project in
their cluster, and the last academic day of the four weeks is
Presentation Day, when the students share the results of their
research. Each cluster handles the research projects in its own
way. In Cluster 7, there were a variety of pre-designed research
projects. A project advisor worked in collaboration with two to three
students to collect, analyze, and interpret data on a specific
topic. Cluster 7 had typically allotted the majority of the last two
weeks to project time, totaling 20 to 30 hours. Our emphasis and
devotion to the small-group research projects was one of the most
unique and successful aspects of Cluster 7. The high staff-to-student
ratio of Cluster 7, and all that it enabled, was largely made possible
through our relationship with the Center for Adaptive Optics
Professional Development Program (CfAO PDP).

Cluster 7 developed alongside and over about the same time period as
the education programs of the CfAO, a National Science Foundation
science and technology center. The CfAO's extensive education and
outreach program is largely driven by its PDP \citep[][also see Hunter
et al., this volume]{PDPdesc}. Through the PDP, scientists, engineers,
and\slash or educators learn inquiry-based teaching techniques and
issues of diversity and equity in the sciences. PDP participants
practice teaching in a range of venues; COSMOS Cluster 7 has often
served as a teaching laboratory for PDP participants. All of the lead
instructors and developers of this course have been participants, and
in addition, we recruited teaching assistants (a.k.a.\ project
advisors) and facilitators for hands-on and\slash or
inquiry activities (see \S\ref{sec.currcomp}) from the PDP community.

Cluster 7 provided an opportunity to reach out to students lower on
the ``leaky science pipeline'' \citep{atkinetal02,heri-attrition} and
encourage and prepare them for college. With this aim in mind, we
recruited and selected students as follows. During the school year,
Cluster 7 instructors and\slash or PDP staff did targeted recruitment,
giving presentations at partner high schools. The partner schools
typically had large numbers of: minorities under-represented in
science; English-language learners; free\slash reduced lunch
recipients; and\slash or (potential) first-generation college-bound
students. COSMOS then provided our cluster with approximately 40
applications of students who passed initial selection criteria and
ranked our cluster as their first or second choice and\slash or
applied as a result of our targeted recruitment. Typically, Cluster 7
instructors performed a holistic review but gave weight to teacher
recommendations and achievement potential despite lesser opportunities
or economic disadvantage. We aimed to give opportunities to students
who may not have attended such a program previously. In addition,
while applicants ranged over four grades, we prioritized the students
entering 10\textsuperscript{th} and 11\textsuperscript{th} grades, the
age where we were best able to affect their college application
pathways. We ranked the applications, which may also have been highly
ranked by another cluster, and then the campus COSMOS office placed
the highest-ranked students into their preferred cluster until all
clusters had 15 to 18 students.

The astronomy course partnered with a vision science course from 2001
to 2006 and a mixed biology course in 2007. This paper focuses only on
the astronomy course.  In \S\ref{sec.currcomp}, we will briefly
describe the various lectures, hands-on activities, and inquiries used
over the years.  We will refer to these components as we outline and
explain the four-week schedule from 2007 in \S\ref{sec.schedule}. How
this specific schedule can be rearranged to accommodate other
considerations will be described in \S\ref{sec.altsched}. Final
thoughts will be discussed in \S\ref{sec.disc}. Other documentation,
e.g., presentations, handouts, etc., from Cluster 7 (2005--2007) can be
found
online.\footnote{\url{http://space.mit.edu/~kcooksey/teaching.html}.}

\section{Curriculum Components}\label{sec.currcomp}

Before we can discuss the rationale behind the astronomy course as a
whole, we need to briefly describe all the lectures, hands-on
activities, and inquiries used in the course over the years. In Table
\ref{tab.currcomp}, we give the title, approximate duration, and basic
purpose of the curriculum components used over the years.  They are
listed, roughly, in the order historically used. In the tables, the
components are described only in terms of their content and
methodology. The motivation for why the components were designed as
such will be explained in \S\ref{sec.schedule} and
\S\ref{sec.altsched}.

\begin{table}[!htb]\small
\caption{Curriculum Components\label{tab.currcomp}}
\smallskip
\begin{center}
\begin{tabular}{p{1.5in}ccp{2.5in}}
\tableline
\noalign{\smallskip}
\multicolumn{1}{c}{\bf Title} & \multicolumn{2}{c}{\bf Length} &
\multicolumn{1}{c}{\bf Brief Description} \\
 & hrs & mins & \\
\tableline
\noalign{\smallskip}
\multicolumn{4}{c}{\bf Lectures} \\
\tableline
\noalign{\smallskip}
Overview \& Introduction & -- & 45 & 
History of cluster and affiliation with CfAO; course topics, schedule, and
expectations. \\
Our Place in the Universe & 2 & 45 & 
Taxonomy of Universe---solar system to super-clusters of galaxies; 
astronomy jargon. \\
Telescopes & 1 & -- & 
Refracting and reflecting
telescopes; telescope size, sensitivity, and
resolution. \\
CCDs & 1 & -- & 
How charge-coupled devices (CCDs) work and are used
in astronomical observing.  \\
Astrophysics I: Cosmology & 1 & 45 & 
Formation,
evolution, and projected future of Universe. \\
Adaptive Optics & 1 & -- & 
How adaptive optics (AO) systems work
and are used in observing. \\
Astrobiology & 1& 30 & 
Short video about variety of life
on Earth; discussion on definition of life and habitability of planets.  \\
Astrophysics II: Stars & 1 & 45 & 
Formation and
evolution of stars of different mass. \\
\tableline
\noalign{\smallskip}
\multicolumn{4}{c}{\bf Hands-on Activities} \\
\tableline
\noalign{\smallskip}
  Telescopes & -- & 20 & 
  Explore properties of refracting telescopes
  with hand-built cardboard telescopes. \\
  Telescopes \& Optics & 3 & -- & 
  Explore telescope optics
  with cardboard telescopes, lenses, mirrors, and ray boxes. \\ 
  Human CCD & -- & 30 & 
  Mimic CCD functions (from receiving
  data to reading it out) with students with buckets as
  pixels and confetti as photons\slash electrons. \\
  \raggedright{Lick Observatory Field~Trip} & 9 & -- & 
  Tour astronomical observatory, attend history lecture, and observe with
  36'' refractor. \\
  \raggedright{Color, Light, \& Spectra (CLS)} & 1 & 30 & 
  Learn about
  continuous, absorption, and emission spectra with spectrographs and
  continuous and emission sources. \\
  Night-Sky Observing & 2 & -- & 
  Observe objects with small
  telescopes and naked eye at UCSC. \\
  \raggedright{Remote Observing (each~group)} & 1 & -- & 
  Observe objects with Lick Observatory 
  1-m telescope from UCSC, for small-group research projects. \\
\tableline
\noalign{\smallskip}
\multicolumn{4}{c}{\bf Inquiries} \\
\tableline
\noalign{\smallskip}
Optics Inquiry & 5 & 45 & Investigate properties of light and
simple optics \citetext{see Raschke et al., this volume}. \\
\raggedright{Color, Light, \& Spectra (CLS) Inquiry} & 5 & 45 & Learn about
continuous, absorption, and emission spectra with spectrographs and
continuous and emission sources.\\
\tableline
\end{tabular}
\end{center}
\end{table}

Three overarching themes of the course were: (i) all astronomers have
to study is light; (ii) scientific investigation is an iterative,
multi-step process; and (iii) the students are capable of scientific
thinking and inquiry. The first theme drove the content taught,
namely: telescopes and instrumentation; physical properties of light;
and current understanding of the Universe. We designed activities that
allowed the students to practice scientific processes (second
theme). By engaging them in doing science, we aimed to have the
students prove to themselves that they could think analytically and
pursue scientific careers if they chose (third theme).

Lectures, hands-on activities, and inquiries were the tools that we
used for any given pedagogical need.  In general, the lectures were
slide presentations. This mode of teaching was the one to which the
students were most accustomed. We would engage the students by
soliciting questions and answers, taking breaks, and integrating the
hands-on activities. In all lectures, there was a minimum of content
that needed to be covered and getting ``side-tracked'' by the
students' interests was completely acceptable.  While potentially
open-ended, the hands-on activities were typically structured around
clear instructions and represented another comfortable endeavor for
the students. On the other hand, each inquiry activity was designed to
subtly guide the students through a full, self-motivated scientific
investigation, from defining a question to sharing results. Raschke et
al.\ \citetext{this volume} describes in detail an inquiry used by
Cluster 7. Students were least likely to be previously familiar with
the inquiry style of learning, and this was taken into account in
designing, scheduling, and implementing the inquiry.

\section{Example Schedule: 2007}\label{sec.schedule}

\begin{figure}[!ht]
\plotfiddle{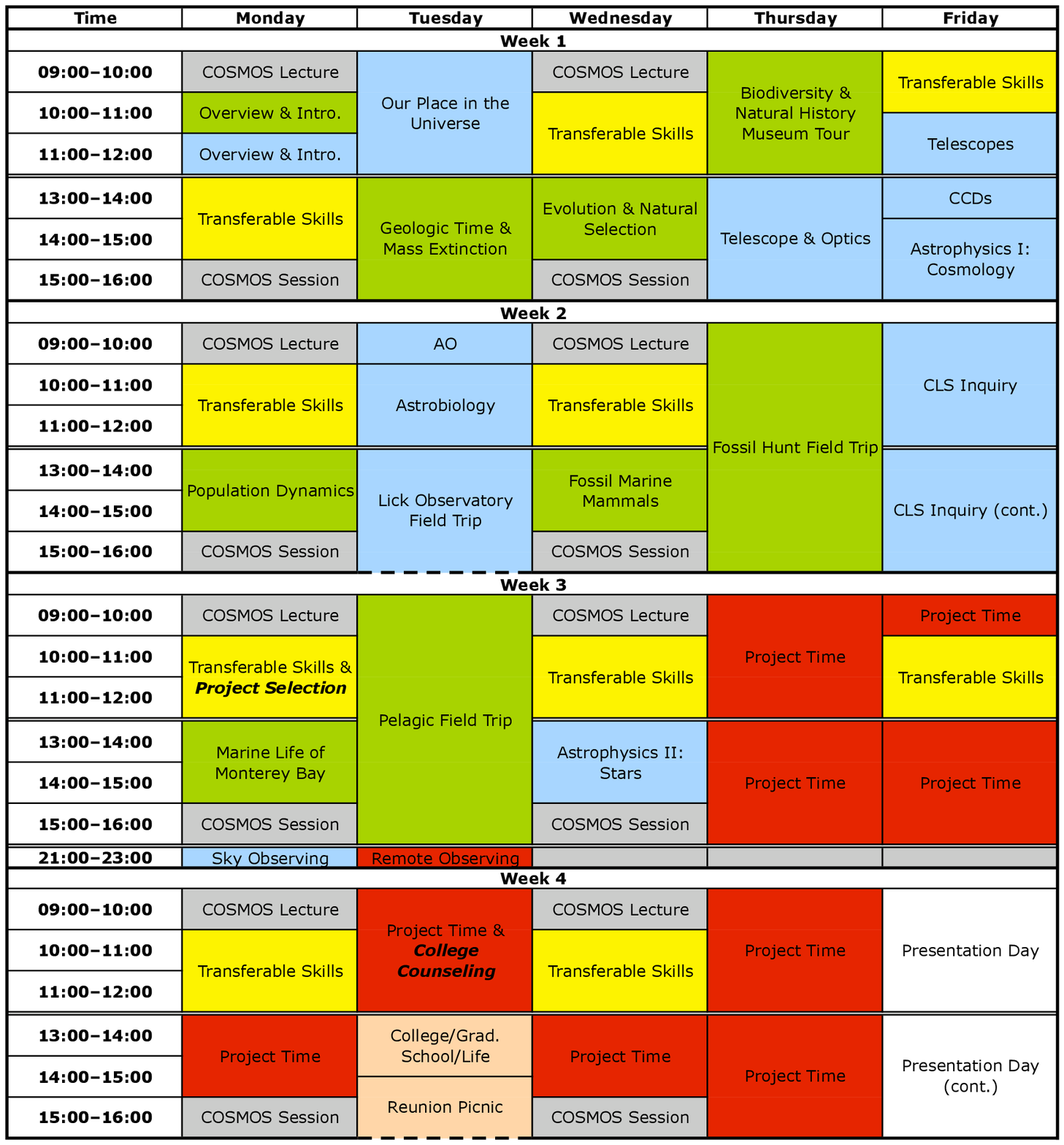}{13.9cm}{0}{75}{75}{-228}{-104}
\caption{Schematic of ``Cluster 7: Stars and Cells'' 2007
  schedule. The astronomy course components (\emph{blue}) are
  described in Table \ref{tab.currcomp}; the ``Remote Observing''
  activity on the third Tuesday is shaded as small-group research
  project time (\emph{red}). Research time blocks are labeled
  generically except for when there was college counseling (fourth
  Tuesday). Similarly, Transferable Skills sessions are not described
  (\emph{yellow}) except for the project selection event (third
  Monday). The partner course was ecology and evolutionary biology and
  paleontology (\emph{green}). COSMOS-wide activities are shown
  (\emph{gray}) though the lunch hours and evening break are just
  thick (\emph{gray}) lines. Events that extended beyond the time
  shown are indicated by dashed lines. In the last week, part of an
  afternoon was for the whole Cluster 7 (\emph{tan}). Presentation Day
  (\emph{not shaded}) included Cluster 7 and two partner
  clusters.\label{fig.schedule}}
\end{figure}

In 2007, ``Cluster 7: Stars and Cells'' featured our astronomy
course and a partner biology course. In this section, we
detail the astronomy course curriculum and explain the rationale. The
four-week schedule is outlined schematically in Figure \ref{fig.schedule}.

There were two perspectives on the goals of the astronomy course: the
students' and the instructors'. The students chose Cluster 7 to learn
about astronomy and\slash or biology; their goal was
content-oriented. The astronomy instructors wanted to develop the
students' abilities to think scientifically, which would benefit them
in any career, though we emphasized science careers; our goals were
process- and motivation-oriented. We engaged the students in the
processes of doing science in the content area(s) in which they were
interested; they had to learn specific content in order to
successfully advance throughout the course and to succeed in their
small-group research project. As mentioned in \S\ref{sec.intro}, all
COSMOS students undertake a research project, and Cluster 7 has
emphasized this (by devoting time and resources---essentially the last
two weeks of the program) more than other clusters. In 2007, the
projects included Variable Stars,\footnote{Documentation for the 2004
  version of the Variable Stars project can be found at
  \url{http://space.mit.edu/~kcooksey/COSMOS/varstars04.html}.\label{fn.varstars}}
Galaxy Morphologies, and Astrobiology.\footnote{For details about the
  2007 Astrobiology project, see Quan et al., \citetext{this volume}.}
These projects drew heavily on previous course content.  (Other
projects developed for other years of Cluster 7 include Globular and
Open Star Clusters\footnote{More details about the 2005 versions of the
  Globular and Open Clusters projects can be found at
  \url{http://www.ucolick.org/~kirsten/COSMOS05/index.html.}\label{fn.clusters}}
and Planetary Nebulae.)  The research projects were a synthesis of
content and processes from the astronomy course.

Since part of the students' final success was giving an outstanding
presentation on their research projects, we wove into the
course opportunities for the students to practice giving talks. They
gave short presentations on a recent Astronomy Picture of the
Day.\footnote{\url{http://apod.nasa.gov/apod/}} Then their peers gave
specific compliments and critiques. Also, our inquiry investigations
required short presentations; the transferable skills course and
partner biology course also incorporated practice presentations.

In 2007, we had 17 students from 13 different schools. The importance
of ice-breakers, bonding, setting the right initial tone, etc., cannot
be overstated. On Opening Day, the whole cluster---instructors,
project advisors, teacher fellow, resident assistants, and
students---gathered and began getting to know each other.

The first astronomy ``lecture'' was a simple introduction to Cluster 7
and an over\-view of the next four weeks. This was the first formal
opportunity to influence how the students \emph{should} participate in
class. We introduced one of the main course themes: all astronomers
have to study is light; we used this as a vehicle to encourage
questioning, participation, and risk-taking. In this case, we showed a
full-color astronomical image and had students make observations and
ask questions about the image. We shared what astronomers observed and
how they interpreted the image.  The ``Overview and Introduction''
lecture thus introduced content and processes and set expectations for
the class: be curious, ask questions, and be respectful of others.

The first content-full lecture was ``Our Place in the Universe,''
where we explored the taxonomy of the Universe from the solar system
to super-clusters of galaxies. This was the only lecture where
galaxies were discussed, and we set the seed of interest for the
Galaxy Morphologies project here.

To address the overarching theme that ``all astronomers have to study
is light,'' the course placed importance on the tools of astronomy:
telescopes, charge-coupled devices (CCDs), and adaptive optics. Since
it was early in the course, we designed a guided activity
(``Telescopes \& Optics'') where the students introduced themselves to
simple refracting telescopes and how optics work. We emphasized that
the students were to explore what interested them with the materials
given. The process of understanding a multi-component, multi-property
system (refracting telescope) was divided into a few steps, within
which there was freedom to come to one's own understanding. After
about an hour of small-group investigation, the students informally
shared with the class their understanding of lenses, light, and
telescopes.

The day after, there was a short traditional lecture,``Telescopes.''
We reviewed what the students taught themselves and each other the
previous day. Then we built upon that to teach how reflecting
telescopes work. The formal presentation of ray tracing was intended
to codify the students' understanding of how lenses and mirrors bend
light and form images. Some students preferred the lecture format, so
addressing this material in two different ways helped assure that
expertise was distributed equitably.

Once the students learned how astronomers collect light with
telescopes, the next step was to teach how data are collected.  The
next lecture was ``CCDs,'' which included the ``Human CCD''
activity. Since verbally explaining how CCDs function was difficult,
we had the students work as a group to mimic a CCD receiving photons
and reading it out (``bucket brigade'' style) to reconstruct the
image. The bucket and confetti ``Human CCD'' was a robust analogy that
allowed us to run through a variety of properties of CCDs. It is also
a very fun activity!

The ``Astrophysics I: Cosmology'' lecture\slash discussion
followed. It was an ideal topic to break the flow.  This was about the
end of the first week, and the students were generally tired, but the
topic of the formation and evolution of the Universe was very
enticing. The ``lecture'' followed their questions to the topics that
interested them.

The two most important events of the astronomy course occurred in the
second week: the observatory field trip and the inquiry
activity. Since an adaptive optics (AO) system is part of the
observatory tour, prior to the field trip was the ``Adaptive Optics''
lecture. After this lecture, the astrobiology project advisors led the
``Astrobiology'' class discussion, beginning with a compelling video
on the diversity of life on Earth, including organisms that live in
extreme environments (see Quan et al., this volume). From this
starter, the class together defined life and tied extremophiles to the
possibility of life on other planets. The students had read an article
explicitly on extremophiles to bolster their knowledge of the
topic. This lecture planted the seeds for students' interest in the
Astrobiology project, similar to what ``Our Place in the Universe'' did
for the Galaxy Morphologies project.

The field trip to Lick Observatory on Mt. Hamilton consisted of: a
long, winding drive, which has always proven to be valuable
unstructured time between instructors and students; a tour of the many
telescopes on the mountain; a picnic dinner; a history talk about the
Observatory; and observing the sky by eye and with a telescope. On the
field trip, a hint was planted for the Variable Stars
project.\textsuperscript{\ref{fn.varstars}} Extrasolar planets are a
main research theme at Lick; one detection method is the decrease in
the brightness of the host star when the planet eclipses it. The
beginning of the Variable Stars project relies on the students
remembering this (or the project advisor facilitating the students
remembering). On the field trip the students also visited and learned
about the 1-m telescope that the astronomy research projects used to
collect data.

In 2007, the showcase inquiry was the ``Color, Light, \&
Spectra (CLS) Inquiry.'' This inquiry, originally created for
undergraduates, was re-designed for high school students by four CfAO
PDP participants; three facilitated the inquiry. The main content
goals were: light is composed of all colors, and there are three kinds
of spectra. The facilitators (instructors) intertwined this content with scientific
reasoning skills in the inquiry, because the students would utilize
both during research project time. All of the projects relied on the
students using the different colors of their astronomical objects to
gain understanding of the underlying physics.  We also wanted the
students to complete an investigation before they moved into project
time. Here again was an opportunity to practice giving a presentation.

The first day of the third week was project selection day, when the
research projects were presented and students ranked their
selections. The instructors sorted the students into groups largely
based on the students' rankings but also with consideration for the
small-group dynamics, an easily-overlooked point that was very
important. Since the instructors handled the sorting, they ultimately
decided who might make good groups without openly exposing this to the
students. This was in contrast to an earlier lottery-style assignment
process, wherein instructors must either abide by the students'
choices or make an obvious (and potentially regrettable) intervention
on-the-fly.

We scheduled an optional night-sky observing excursion for that
evening.  On the following evening, the astronomy research project
groups, Variable Stars and Galaxy Morphologies, met for ``Remote
Observing'' using the 1-m telescope on Mt. Hamilton, from the CfAO
conference room. The students were in communication with the
astronomer (their Lick tour guide) at the Observatory, who
assisted in the observations. The remote observing capabilities gave
students the identical experience of any astronomer, a point which we
emphasized to reinforce the students-as-scientists theme.

The last astronomy lecture was ``Astrophysics II: Stars,'' which built
heavily on the CLS Inquiry content. The main points were: short-lived
massive stars are hot, bright, and blue, and less massive stars are
the opposite (cool, faint, and red). The Galaxy Morphologies project
relied and expanded on these principles.

For the last week and a half, the students focused largely on their
research projects. However, our cluster also planned several
bigger-picture elements to break up this research time and address
extremely important topics. Early in the last week, we organized a
UCSC admissions counselor to review and discuss the students'
transcripts. The counselor suggested what classes to take and grades
to aim for, in order for the student to qualify for acceptance to
UC. This individual attention and analysis was not necessarily the
sort that our students had received or could receive at their schools
(see \S\ref{sec.intro} for details on student selection).  The
students met with the counselor in groups of two to three, by
grade. This divided the research groups, so the project advisors had
to account for this. Any students who felt reserved about their
transcripts met with the counselor individually.

The same day Cluster 7 met for a ``college\slash graduate
school\slash life'' overview and panel discussion. The overview part
was a summary of the steps from high school to graduate school and
suggestions for along the way (e.g., where to find
scholarships). Then, we made four groups of students and staff and
spent one and a half hours fielding questions and sharing
experiences. We mixed up the students in the groups half way through
to increase the interactions. Since many of our students would be the
first college students in their families, they had many questions,
worries, and uncertainties about pathways to, through, and from
college. After all the talk about the future, the students had a
chance to ``meet the future'' via the annual Cluster 7 reunion picnic
on the beach. We invited all previous Cluster 7 students and staff to
share how the COSMOS experience played out afterward. 

The last academic day was Presentation Day. Cluster 7 has always
performed exceptionally well on Presentation Day because the project
advisors devoted significant time to working on the presentations with
the students. As the students developed and practiced their group
presentations, we were assessing what they understood in order to
teach any final points and to evaluate our success as teachers.

\section{Motivating Alternative Schedules}\label{sec.altsched}

From 2001 to 2006, Cluster 7 was known as ``Stars, Sight, and
Science'' and was a pairing of astronomy and vision science
courses. The original astronomy course was designed in conjunction
with the vision science course, and we made choices about the
astronomy components in light of what would also be beneficial for
both. The 2007 schedule largely contained the same astronomy
components in the same order. Changes were made to accommodate the new
partner course and incorporate newly designed activities. Here we
rationalize some of the key differences in the schedules.

\begin{figure}[!ht]
\plotfiddle{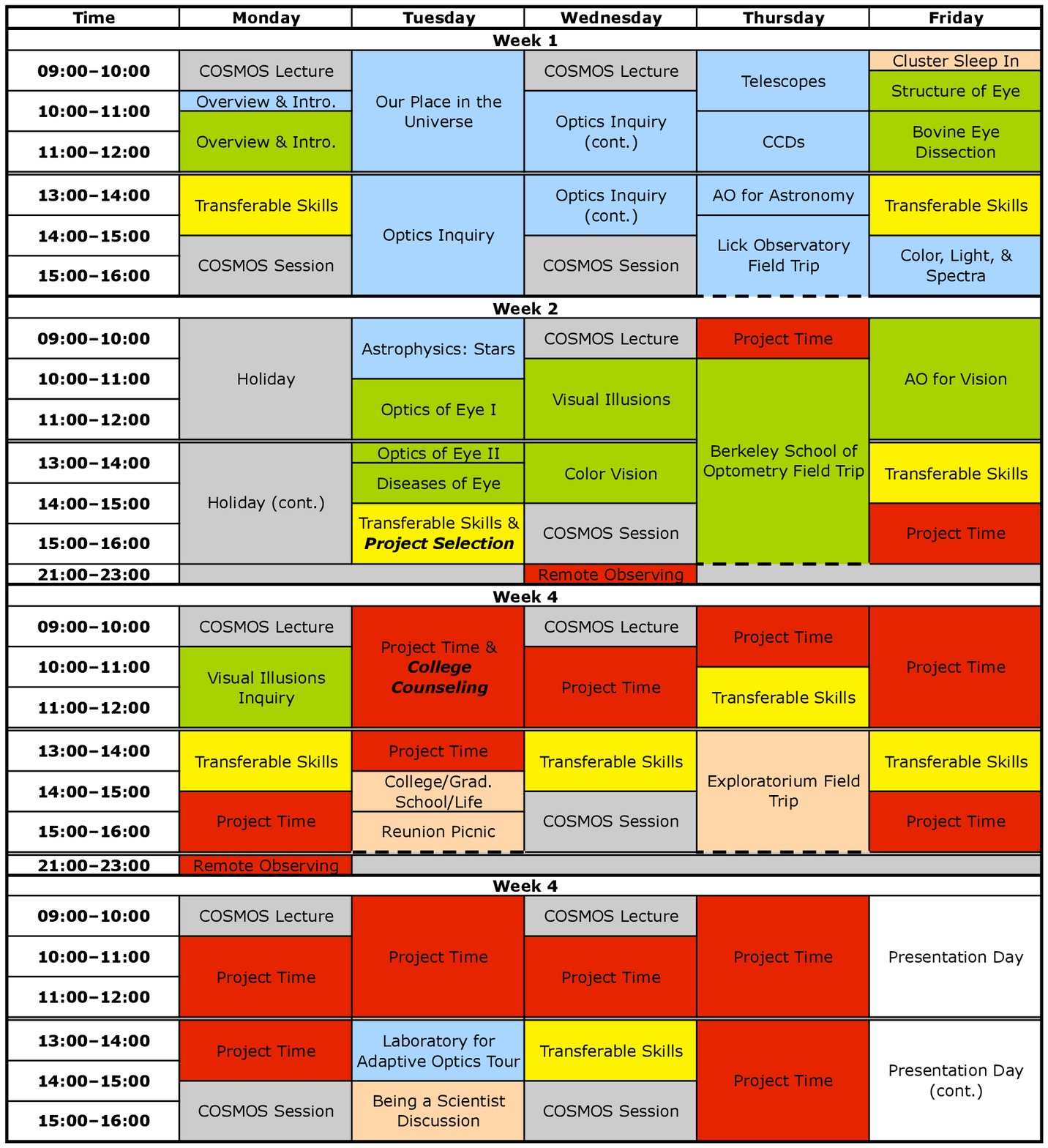}{14.2cm}{0}{75}{75}{-228}{-101}
\caption{Schematic of ``Cluster 7: Stars, Sight, and Science'' 2005
  schedule. The color coding is the same as used in the 2007 schematic
  (Figure \ref{fig.schedule}), except the 2005 partner course was
  vision science (\emph{green}). Though this schedule had a fairly
  distinct ``astronomy course week'' (first) and ``vision science
  course week'' (second), the whole Cluster was more cohesive due to
  the extensive collaboration between the astronomy and vision science
  instructors, all from the PDP community, during the curriculum
  design. In addition, there was more topical cohesion since both
  courses taught about optics, color, and adaptive
  optics. \label{fig.altsched}}
\end{figure}

The main differences in the ``Stars, Sight, and Science'' schedule
were as follows: there were no ``Cosmology'' or ``Astrobiology''
lectures\slash discussions; we had an ``Optics Inquiry;'' we used the
``Telescopes'' activity in the ``Telescopes'' lecture; and CLS was a
hands-on activity instead of an inquiry (see Figure \ref{fig.altsched}).

There was more time for astronomy lectures in the 2007 schedule since
there was one less field trip and less research project time. There
have always been three main field trips; in 2007, they were Lick
Observatory, a pelagic boat tour, and a fossil hunt. In the years of
``Stars, Sight, and Science,'' the last two trips were to the Berkeley
School of Optometry and a science museum, the Exploratorium. A fourth
field trip, to the Mystery Spot, tied in with the visual illusions
taught.  During the field trips, students were introduced to
professionals who were using and\slash or doing what the students were
learning and\slash or doing. They also provided important unstructured
time for student-staff interactions, but the loss of one trip was
necessary due to budget constraints.  To compensate, we had a picnic
on campus and a short tour of the UCSC Lab for Adaptive Optics.  We
learned that the picnic was not a good choice, since the students
preferred eating in the dining hall where they caught up with their
friends from other clusters. This was a lesson in coordinating with
the larger program. The Lab tour had the benefit of allowing the
students to see a simple AO system \citetext{Harrington et al., this
  volume}.

There was less project time in 2007 because the other course elected
to cover more material than the previous vision science course.
Supporting the needs of the research projects was a priority and
affected the final schedule, including an entirely new
``Astrobiology'' lecture in 2007.  That year there were five students
participating in the Astrobiology project. This project was designed
specifically for the 2007 Cluster 7 \citetext{Quan et al., this
  volume}; it acted as unifying topic to the two courses.

When paired with vision science, we used the ``Optics Inquiry''
instead of the CLS Inquiry since a good understanding of optics would
support the astronomy and vision science content. For 2007, we
modified the Optics Inquiry into the ``Telescope \& Optics'' activity
to convey similar content and to retain elements of the students
investigating on their own. The color, light, and spectra content was
more fundamental to all of the astronomy research projects, so when we
were not paired with vision science in 2007, we chose the CLS Inquiry
as the only inquiry activity. With even more astronomy projects (the
two Star Clusters projects\textsuperscript{\ref{fn.clusters}} and Planetary
Nebulae), the CLS content would be even more fundamental and drawn
upon in project time.

Another substantial difference between the 2007 schedule and prior
years' was the distribution of the astronomy and other courses' time
blocks.  Setting the schedule was an iterative process. Important
scheduling constraints came from the COSMOS-wide office, the staff's
availability, the field trips, and time-sensitive observing
(i.e., Variable Stars project\textsuperscript{\ref{fn.varstars}}). But the
observing had to come after the students learned about astronomical
observing. The astronomy project groups also needed the data early in
their projects, so observing could not be arbitrarily late in the four
weeks.  When paired with vision science, the astronomy course was scheduled to
fit mostly in the first week, leaving a vision science block for its non-UCSC
instructor to schedule travel around. 

There were down-sides to isolating the astronomy course to the first
week and the vision science course to the second. By project selection
time, the students had more vision science on their minds. We also
taught the astronomy content at an accelerated rate to make a basis
for the Lick Observatory field trip.  The more interwoven schedule
described in \S\ref{sec.schedule} was useful in preventing burn
out. The switching between instructors and topics gave the students
more times to let what they learned settle. The trend in scheduling
Cluster 7 from 2005 to 2007 was to compact project time in order to
minimize the time the project advisors were involved because they were
typically busy graduate students.

However, with the partnership of astronomy and vision science came
significant benefits: extensive collaboration during curriculum
design; cohesive teaching practices; and shared content. As mentioned
previously, Cluster 7 was originally developed by astronomers and
vision scientists, educated and trained by the CfAO PDP. In addition,
optics and properties of light (e.g., color) are fundamental content
for both sciences, and adaptive optics is a cutting-edge research
field in astronomy and vision science, which is why the two fall under
the purview of the CfAO.

The overall flow of Cluster 7 was similar to an inquiry.  Dividing
into small research groups for the last two weeks of the program had
everyone focusing on their own investigations. In 2006, we designed a
cluster ``synthesis'' for astronomy. We facilitated small groups in
organizing their astronomy content into a concept map. We wanted the
students to think about the big picture and place what they had
learned in it. This activity also allowed us to assess what the
students had retained from different aspects of the course and how
they prioritized the information.

\section{Discussion}\label{sec.disc}

The curriculum components were a mix of slide-based lectures, hands-on
activities, and inquiries. The content taught with any given component
could be conveyed, at least in part, by another format. Indeed, the
topic of telescopes was taught through a combination of lecture,
hands-on activity, and inquiry over the years, as needed to meet
changing goals.

Mental and physical fatigue is a concern for a long, intensive program
like COSMOS. We tried to build in breaks by switching between formats.
We also scheduled less intensive, student-driven lectures amidst the
more formal, content-driven lectures. Field trips are partially
mental breaks but they are physically draining. Also, the students do
not necessarily like missing meal times when they could visit with
their friends from other clusters.  In the earlier years of the
cluster, under the umbrella of the CfAO, the astronomy and vision
science courses cooperated more on content, process, and motivation
themes. Due to scheduling constraints, the two courses were separated,
and this made the cluster more intense and fatiguing. The scheduling
of ``Stars and Cells'' was less fatiguing, and learning benefited from
spreading ideas apart.

The curriculum components of the astronomy course are readily
integrable into other astronomy courses. The hands-on and inquiry
activities are especially useful tools that motivate the students to
learn on their own and gain their own understanding. As was done with
the optics inquiry and the telescope and optics activities, the longer
activities can be scaled down in time, at the expense of some goals.

In talking with previous Cluster 7 students, via email or when they
end up attending UCSC for college, all emphasize the companionship of
COSMOS (residential life) common to all clusters. Our cluster's
extensive field trips (more than other clusters), and the small-group
research projects---each of which benefit enormously from excellent
student-to-instructor ratios--- come up next. The course's connection
with the PDP community has been an integral part of its success.

\acknowledgements We would like to thank: the support of Lisa Hunter
and the education theme within the NSF CfAO (AST--9876783) which
fostered the early development of this course; Malika Moutawakkil
Bell, Hilary O'Bryan, and Liz Espinoza, who coordinated and supported
the course for many years; Carrol Moran (former UCSC Educational
Partnership Center director), Jamie Alonzo and Nafeesa Owens (former
directors of COSMOS at UCSC), and Dan Aldrich (with the UC Office of
the President) who supported this course's innovations within COSMOS;
our Lick Observatory guru Ellie Gates; John Martin, Gary Martindale,
and Jeff Sweet, our exceptional teacher fellows; Julia Kregenow,
Jeremy Wertheimer, and Diane Wong, 2007 inquiry designers and
facilitators; the UCSC Department of Astronomy for supporting their
graduate and postdoctoral students participating in COSMOS; and years
of research project advisors: Phil Choi, Laura Chomiuk, Kristel
Dorighi, Emily Freeland, Marla Geha, Javiera Guedes, Kirsten Howley,
Jess Johnson, Patrik Jonsson, Evan Kirby, Nick Konidaris, David Lai,
Jennifer Lotz, Jason Melbourne, Anne Metevier, Ryan Montgomery, Stuart
Norton, Giovanna Pugliese, Shannon Patel, Tiffani Quan, Yancey
Quinones, Kate Rubin, and Anouk Shambrook. For some of this work,
author S.~Seagroves was supported by the NSF Center for Informal
Learning and Schools (DRL--0119787).

\bibliographystyle{asp2010}
\bibliography{kcooksey_Cluster7}

\end{document}